\documentclass[10pt,journal,compsoc]{IEEEtran}

\usepackage{url}
\usepackage{algorithm}
\usepackage{array}
\usepackage{cite}
\usepackage{graphicx}
\usepackage{times}

\begin{document}


\title{Heterogeneous Data Access Model for Concurrency Control and Methods to Deal with High Data Contention}  

\author{Alexander Thomasian,~\IEEEmembership{Life~Fellow,~IEEE}
\IEEEcompsocitemizethanks{\IEEEcompsocthanksitem 
A. Thomasian is with Thomasian \& Associates, Pleasantville, NY.\protect\\
E-mail: alexthomasian@gmail.com}
\thanks{Manuscript completed November 7, 2022, revised April 1, 2024.}}
\maketitle


\begin{abstract}
OLTP has stringent performance requirements defined by Service Level Agreements.
Transaction response time is used to determine the maximum throughout in benchmarks. 
Capacity planning tools for OLTP performance are based on queueing network models for hardware resources 
and database lock contention has a secondary effect on performance.
With ever increasing levels of e-commerce and surges in OLTP traffic  
we discuss the need for studies of database workloads to develop more realistic lock/latch contention models. 
Predictive formulas to model increased load leading to thrashing 
for txns with identical and nonidentical steps are presented.
We review concurrency control methods to reduce the level of lock/data conflicts in high contention environments.  
\end{abstract}

\begin{IEEEkeywords}
Transaction processing, concurrency control, two-phase locking, lock contention, 
deadlocks, performance analysis, thrashing, optimistic concurrency control,
heterogeneous data access model, queueing network model.
\end{IEEEkeywords}

\section{Introduction}\label{sec:intro}

{\it Data Base Management Systems - DBMS}s are an extension of file systems,
where multiple interrelated files are stored together.
Early network and hierarchical DBMSs were linked lists on disks, 
which required complex programming to navigate the lists.
Unlike files which are used in batch systems, 
most DBMSs are used in {\it OnLine Transaction Processing - OLTP} environments,
which to ensure database integrity and transaction - txn correctness require {\it Concurrency Control - CC}.

Relational databases use keys to uniquely identify rows or records in tables
and foreign keys to link rows, e.g., students to departments,
so that department information need not be repeated in student records.
Logical links replace physical links (disk addresses),
which results in increased processing cost for relational queries.
For frequent queries denormalization can be used to avoid costly joins \cite{RaGe02}.
The SQL programming language developed for manipulating relational databases
results in tremendous increase in programming productivity,
making the increased cost of hardware resource consumption worthwhile. 
The efficiency of execution plans to process SQL programs depends 
on the quality of query optimizers \cite{RaGe02}.
For a given logical database design a good physical DB design is important for the sake of performance,
e.g., given a list of queries determine indices.

CC has two basic locking modes: Shared - S and eXclusive X, 
where only S locks are compatible with each other. 
Deadlocks arising due to lock upgrades from S to X are dealt with by introducing Update - U locks,
which are only compatible with S-locks.
According to {\it two-Phase Locking - 2PL} txns 
which acquire locks during their execution do not release locks until all locks have been acquired,
while strict 2PL facilitates recovery by releasing locks when txn completes and commits.
Txn recovery requires logging after and before images of modified data \cite{GrRe93},\cite{RaGe02},\cite{BeNe09}.

As part of the development of System R relational DBMS a well-defined hierarchy 
of locking modes was developed and is adopted by other relational DBMSs \cite{GrRe93},\cite{RaGe02},\cite{BeNe09}.
Figure 8.5 in \cite{GrRe93} depicts a popular lock with a FCFS queue.                 \newline
Granted locks (IS, IX, IS, IS, IS), Waiting locks (S, IS, IX, IS, IS)                 \newline
IX/IS are intention locks implying that the table holds records locked in X or S mode. 
When all X-locks are released then the IX lock is released too:                       \newline
Granted locks (IS, IS, IS, IS, S, IS), Waiting locks (IX, IS, IS).

There have been numerous abstract analytic and simulation studies of database lock/data contention, 
such as S and X-locks lock requests uniformly distributed over all database objects \cite{Thom96a,Thom98}. 
There have been few analyses of lock traces to characterize locking behavior \cite{YDR+85},\cite{SiSm97}.
Lock and latch contention is also measured, 
where latches are a low level serialization mechanism which are released 
as soon as the processing on the locked object is completed. 
The case against latch-free algorithms in a multicore systems is made in \cite{FaAb17}.
Studies of database lock contention are reported in \cite{Syba11} in the context of Sybase DBMS \cite{Syba11}
and \cite{Otta18} in the context in IBM's Db2 in the Appendix.  \newline
https://www.cs.umd.edu/~abadi/papers/determinism-vldb10.pdf

According to Chapter 22 in \cite{Krog20} locking 
may cause severe performance degradation in the popular MySQL databases. 
There are four categories of lock issues: 
flush locks (making tables readonly), metadata locks, record-level locks, and deadlocks. 

In addition to locking {\it Optimistic CC - OCC} \cite{KuRo81},\cite{FrRT92},
also described as "Improved Conflict Resolution", on p.568) in \cite{RaGe02}
and {\it TimeStamp-Ordering - TSO} method 
were compared via analysis and simulation in \cite{RyuI85}. 
TSO was shown to be outperformed by other two methods although it was not evaluated with 
{\it MultiVersion CC - MVCC} schemes \cite{RaGe02},\cite{WeVo02}. 

The paper is organized as follows. 
Lock contention studies limited to X-locks and S-locks are discussed in Section \ref{sec:locking}. 
The analyses of a single and multiple txn classes accessing a single or multiple {\it Data Base Regions - DBRs} 
are discussed in Section \ref{sec:single} and Section \ref{sec:multiple}, respectively.
Analytic methods to evaluate the performance of txn processing systems  
taking into account lock contention are discussed in Section \ref{sec:degraded}.
Methods to prevent deadlocks and reduce lock contention are discussed in Section \ref{sec:loadcontrol}.
Conclusions are drawn in Section \ref{sec:conclusion}.
A measurement tool for IBM's Db2 is outlined in the Appendix.

\section{Lock Contention Studies}\label{sec:locking}

An early study applied {\it Queueing Network - QN} modeling \cite{LZGS84}
to determine the effect of locking granularity in a concurrent txn processing system \cite{IrLi79}.
The conclusion to adopt a coarse locking granularity may be due to high locking overhead postulated.

The probability of lock conflict $(p_c)$ and deadlock 
in txn processing systems is reported in \cite{GHOK81} and then in \cite{GrRe93}. 
Txns request X-locks uniformly from a database of size $D$,
which is the number of database objects, e.g., pages.
Given that txns consist of $k+1$ steps with equal processing times
and X-locks are requested at the conclusions of first $k$ steps 
and released at the end of $k+1^{st}$ step as the txn commits then the mean number of X-locks 
held by txns is $\bar{k} \approx k/2$. 
Given that each step takes $\bar{s}$ time units dividing the time-space area held locks by txn processing time we have:
\vspace{-1mm}
$$k(k+1)\bar{s}/2 / (k+1)\bar{s} \approx \bar{k}=k/2,$$  
Lock contention can be reduced significantly by deferring lock requests, 
especially to hot objects as much as possible.
Muti-phase processing methods reduce lock holding times,
since lock utilizations are a good indicator of $p_c$.

Given that txns only request X-Locks and the degree of txn concurrency is $M$ 
then a txn requesting a lock may encounter a lock conflict with probability:
\vspace{-2mm}
\begin{eqnarray}\label{eq:pc}
p_c = \frac {\bar{k} (M-1)}{D} \approx \frac{ k (M-1) }{2D}
\end{eqnarray}

The analysis in \cite{GHOK81} leads to the equation on the left for 2-way deadlocks,
but it is argued in \cite{ThRy91} that a 2-way deadlock will occur 
only if the txn requests a lock held by another txn is already blocking it.
The mean waiting time of txns blocked by active txns $(W_1)$
normalized by txn response time is $A=W_1/R \approx 1/3$ for fixed sized txns,
which leads to the modified equation:

\vspace{-2mm}
\begin{eqnarray}
p_{2-way} \approx \frac{(M-1)k^4}{4 D^2},
\hspace{5mm}
P_{2-way}^{modified} \approx \frac{(M-1)k^4}{12 D^2}
\end{eqnarray}

The analysis in \cite{IrLi79} to study the effect of granularity of locking and 
{\it MultiProgramming Level - MPL} is extended in \cite{Thom82}.
where an iterative solution method is developed 
to obtain system performance under high levels of lock contention.
Lock contention was increased by increasing $k$ and beyond a certain point 
the iteration ceases to converge as the system thrashes. 

The analysis of lock conflict and deadlocks in \cite{GHOK81} 
is extended to variable size txns in \cite{ThRy91},
where extensive simulation results for validation are presented.
The effect of deadlocks on performance degradation in txn processing systems 
was taken into account in \cite{Thom82},\cite{RyTh86,RyTh90},
but ignored in \cite{Thom91,Thom92,Thom93}, 
since deadlocks are rare and have a negligible effect on performance degradation.

Considered in \cite{TaGS85} is txn processing with 2PL CC with $k$ lock requests per txn.
It is noted that txn aborts and restarts to resolve deadlocks have a small effect on txn throughout.
The maximum load leading to thrashing is shown to be: $k^2 M /D \approx 1.5$. 

If a fraction $b$ of lock requests are to a fraction $c$ of the database
then the effective database size for uniform accesses 
and if a fraction $s$ of lock requests are shared is given as follows \cite{TaGS85}:
\vspace{-2mm}
\begin{eqnarray}\label{eq:shared}
D_{eff}^{skewed} = D [\frac{b^2}{c}+ \frac{1-b)^2}{1-c}]^{-1} \hspace{5mm}
D_{eff}^{mixed} = D  (1-s^2)^{-1}.
\end{eqnarray} 

A txn requesting an X-lock on an object held by txns with S-locks on the object
generates a deadlock if one or more of those txns is blocked  by it. 
Similar concepts on deciding which txn to abort 
were considered in the context of the CC WDL method discussed below
using the number of requested locks but the number of byte written to 
the log is another measure of txn progress and its priority in deadlock resolution. 
 
\section{Analysis with a Single Txn Class}\label{sec:single}

Given that the lock conflict probability $p_c(\lambda)$ at txn arrival rate arrival $\lambda$ is known, 
then $p_c(\lambda')$ for $\lambda' > \lambda$ can be estimated as follows:

\vspace{-2mm} 
\begin{eqnarray}\label{eq:extrapolate}
p_c (\lambda') = p_c (\lambda) 
\frac{ \lambda' R ( \lambda') }{\lambda R (\lambda) }
\end{eqnarray}
This is because of the variation of the mean number of held by increased mean number of active txns 
$\bar{M}=\lambda R(\lambda) $ according to Little's result \cite{LZGS84} in Eq. (\ref{eq:pc}),
where $R(\cdot)$ is the mean txn response time. 

It is assumed that there has been no change in txn lock request pattern and the database size has not changed,
but in a banking application the increase in $\lambda$ is usually due to an increase in the number of bank customers, 
sometimes due to bank mergers resulting in an increased number of customer records ($D=D_1+D_2$).
The increased level of activity is then by the increased number of customers.

To quantify the effect of increased txn arrival rates on lock contention in an open QN model consider 
a Poisson arrivals at rate $\lambda$ to a computer system with $N=3$ devices: 
CPU and two disks with equal service demands $X_n=X =100, 1 \leq n \leq 3 $ milliseconds - ms.
and utilization factors $\rho_n=\lambda X_n = 1/3, 1 \leq n \leq N$,
The mean txn response time is then:                                              
$$R (\lambda) = \sum_{n=1}^N r_n (\lambda) = 3 \times X/(1-\rho)=450\mbox{ ms.}$$   
Doubling the txn arrival rate: $\lambda'=2 \lambda$,
results in $\rho=2/3$ and $R\lambda') = 900\mbox{  ms.}$
Referring back to Eq. (\ref{eq:extrapolate}) $p_c (\lambda') = 4 p_c (\lambda). $

The maximum {\it Degree of MultiProgramming - DMP} ($M_{max}$) is determined by memory size 
and may limit the maximum throughput determined by the {\it Asymptotic Job Bound - AJB} \cite{LZGS84}. \newline
$\lambda_{max}=1/D_{max} = 1/100 =10$ Txns Per Second - TPS.                                    

For the proposed balanced computer system with $N$ devices 
with $M$ jobs txn throughput is given as \cite{LZGS84}:
$$T(M)=\frac{M}{N+M-1}\frac{1}{X}.$$
The minimum MPL ($M_{min})$ should be large enough to process txns with its current arrival rate ($\rho=2/3$):
\vspace{-2mm}
\begin{eqnarray}\label{eq:minMPL} 
\frac{ M_{max} }{N + M_{max} -1}\frac{1}{X} > \lambda\mbox{  or } 
M_{max}  > \frac{(N -1) \rho }{1-\rho} = 4.  
\end{eqnarray}

In addition to txn throughput it is important meet txn response time requirements,
which is the sum of device residence times and possibly delay in memory queue before activation.
An important consideration is the database buffer hit ratio and also the flush rate,
which is estimated for MySQL databases in \cite{MCJM13}

The discussion can be extended to multiple txn classes to determine the DMP 
and the speed of computer devices to meet response time requirements by SLA.
Txn classes may run in {\it Message Processing Regions - MPRs} assigned to them, 
which restrict their degree of concurrency and hence the lock contention in DBRs accessed by them. 

There is a need to characterize the service demands of txns in utilizing hardware resources 
to analyze the respective QN model using tools such as IBM's {\it Systems management Facilities - SMF} 
which includes {\it Resource Measurement Facility - RMF}.
SMF and RMF measurements  are used as inputs to BEST/1 \cite{Buze78} and MAP \cite{LZGS84} capacity planning tools.
{\it Transaction Processing Council - TPC} has defined multiple benchmarks,
such as order-entry TPC-C OLTP benchmark \cite{tpcc}, which have been used in performance studies,
but introduces a low level of lock contention.

\section{Multiple Txn Classes Accessing a Single or Multiple DBRs}\label{sec:multiple}

In an open system txns in class $i$ or ${\cal C}_i$ arrive with rate $\lambda_i= \mathrm{\Lambda} f_i$,   
where $\mathrm{\Lambda}$ is the total arrival rate and 
${\bf f}=\{f_1,f_2,\ldots,f_I\}$ is the frequency of different classes. 

We first consider multiple txns classes with fixed frequencies accessing a single DBR.  
Given job service demands we compute mean txn response times 
$R_i (\lambda_i)$ and the mean number of txns in class ${\cal C}_i$:
$$\bar{M} (\lambda_i) = \lambda_i R(\lambda_i), 1 \leq i \leq I. $$ 
Given that txns in ${\cal C}_i$ request $k_i$ locks and 
holds $\bar{k}_i$ locks on the average then the mean the mean number of held DBR locks is:
$$\sum_{i=1}^I \bar{k}_i M_i = \sum_{i=1}^I \lambda_i R_i(\lambda_i) \bar{K}_i$$ 
If the txn arrival rates is increased to $\mathrm{\Lambda}' > \mathrm{\Lambda}$ then:

\vspace{-2mm}
\begin{eqnarray}
p_c (\mathrm{\Lambda'}) = p_c (\mathrm{\Lambda})  \frac
{ \sum_{i=1}^I \bar{k}_i {\lambda'}_i R_i ( {\lambda'}_i) }
{ \sum_{i=1}^I \bar{k}_i  \lambda_i   R_i ( \lambda_i ) }.
\end{eqnarray}

In the more realistic {\it Heterogeneous Database Access Model - HDAM}  
txns access multiple DBRs $J > 1$\cite{Thom94,Thom96}.
Txns in ${\cal C}_i, 1 \leq i \leq I$ make $k_{i,j}$ X-lock requests 
to DBR$_j, 1 \leq j \leq J$ holding $\bar{k}_{i,j}$ locks on the average
The probability of lock conflict when requesting locks in DBR$_j$ is:

\vspace{3mm}
\begin{eqnarray}\label{eq:pcj}
p_c^j ({\bf \lambda}) = 
\frac{1}{D_j} 
\sum_{i=1}^I \bar{k}_{i,j} \bar{M}_i 
=
\frac{1}{D_j} 
\sum_{i=1}^I \bar{k}_{i,j} \lambda_i R_i (\lambda_i) 
\end{eqnarray}

As the txn arrival rates varies

\vspace{-3mm}
\begin{eqnarray}\label{eq:extrapolate2j}
p_c^j ( {bf \lambda'} ) = 
p_c^j ( {\bf \lambda} )  
\frac
{ \sum_{i=1}^I \bar{k}_{i,j} {\lambda'}_i R( {\lambda'}_i)   }
{ \sum_{i=1}^I \bar{k}_{i,j} \lambda_i R(\lambda_i)   }
\end{eqnarray}


The fraction of shared locks may vary across txn classes and DBRs. 
Given that txns in ${\cal C}_i$ arriving with txn rate $\lambda_i$ request $k_{i,j}$ locks 
and a fraction of their lock requests to DBR$_j$ are S-locks according to $s_{i,j}$
then the fraction of shared lock request to DBR$_j$ is given as 
the ratio of shared and all locks requested by all txn classes.
$$s_j = \sum_{i=1}^I f_i k_{i,j} s{i,j} / \sum_{i=1}^I f_i k_{i,j}.$$
Eq. (\ref{eq:pcj}) can be rewritten as follows:

\vspace{-2mm}
\begin{eqnarray}\label{eq:pcj2}
p_c^j ({\bf \lambda}) = 
\sum_{i=1}^I 
\frac{\bar{k}_{i,j} \bar{M}_i }
{D_{eff}^{j}} 
= 
\sum_{i=1}^I 
\frac{\bar{k}_{i,j} \lambda_i R(\lambda_i)  }
{D_j/(1-s_j^2)} 
\end{eqnarray}


HDAM with simplifying assumptions was analyzed in \cite{Thom94,Thom96}  
to obtain txn response time degradation due to lock conflicts.
HDAM was adopted in \cite{dPC+10} and 
was applied in the performance study of MySQL in \cite{MCJM13}
with several extensions to the model which were implied but not specified:
(0/1/2) Txn types, DBRs and access pattern were extracted from the MySQL log.
(3) txn resource consumption was estimated, 
(4) infinite resources were assumed for the sake of brevity, 
(5) only exclusive locks, but there is an easy fix (see Eq. (\ref{eq:shared}). 
(6) Txns were assumed to access a fixed number of locks, again for brevity.
In summary: $T=mbox{min}(T_{disk},T_{CPU},T_{lock})$.  


\section{Degradation in Txn Response Time Due to Lock Contention}\label{sec:degraded}

When in a txn processing system with txn arrival rate $\lambda$ 
with txn response times $R(\lambda)$ a fraction $\beta$ of txns are blocked, 
so that the fraction of time txns are active is $1-\beta$. 

\vspace{-2mm}
\begin{eqnarray}
r(\lambda) = R(\lambda)/(1-\beta)\mbox{  or  }R(\lambda) = \frac{ r(\lambda) }{1-\beta}.
\end{eqnarray}

There are $K$ txn classes where the txn class determines the number of requested X-locks,
so that the mean number of requested locks is $K_1 = \sum_{k=1}^K k f_k$.
Given that $\beta = K_1 P_c A$, $P_c$ probability of lock conflict,
and $B= W/R(M)$ the mean waiting time per lock conflict normalized by mean txns response time $(R(M))$.
Assuming that all lock conflicts are with active txns at level 1 
and the normalized delay is $A=W_1 / R(\lambda)$ define $\alpha = K_1 p_c A$ 
where $p_c \approx \lambda R(\lambda) \bar{k} /D$
and setting $a= \lambda K_1 \bar{k} A/D$ we have the following quadratic equation:
\vspace{-2mm}
\begin{eqnarray}\label{eq:RL2}
R(\lambda) = \frac{ r(\lambda)}{1-\alpha}\mbox{  or  } a  R^2(\lambda) - R(\lambda) + r(\lambda) =0.  
\end{eqnarray}

Since $R(\lambda)=0$ when $r (\lambda)=0$ only the negative root of the quadratic is acceptable
and a solution exits for $4a r(\lambda) \leq 1 $

\vspace{-2mm}
\begin{eqnarray}\label{eq:RL3}
R(\lambda)= \frac{1}{2a} \left [ 1- \sqrt{1-4ar(\lambda)} \right]
\end{eqnarray} 

The analysis in \cite{Thom91,Thom93} considers a closed system with $M$ txns a
which allows variable size txns, but ignores deadlocks leads to a cubic equation in $\beta$.
The fraction of blocked txns can be expressed as $\beta = K_1 p_c $,
$K_1$ is the mean number of lock requests per txns,
$p_c$ the probability of lock conflict, 
and $B= W/R(M)$ is the normalized mean waiting time per lock conflict.
The probability of blocking at level $i > 1$ and by an active txn 
at level $i=1$ is given by Eq.~\ref{eq:Pb}
\vspace{-2mm} 
\begin{eqnarray}\label{eq:Pb}
P_b (i) = \beta^i,\hspace{3mm} i \geq 1, \hspace{3mm} p_b(1) =1 -\sum_{i \geq 1}\beta^i.
\end{eqnarray}

So that the mean waiting time at level $i$ and overall waiting time is:

\vspace{-2mm}
\begin{eqnarray}\label{eq:Wi}
W_i \approx (i+0.5) W_1 , i >  1, \hspace{5mm}  W= \sum_{i \geq 1} P_b (i) W_i
\end{eqnarray}

Multiplying both sides of the later equation with $K_1 p_c / R (M)$,
assuming an infinite number of blocking levels. and simplifying:

\vspace{-2mm}
\begin{eqnarray}\label{eq:equal}
\beta^3 - (1.5\alpha+2)\beta^2  + (1.5\alpha +1)\beta -\alpha = 0,
\end{eqnarray}

\begin{figure}[t]
\begin{center}
\includegraphics[scale=0.50,angle=00]{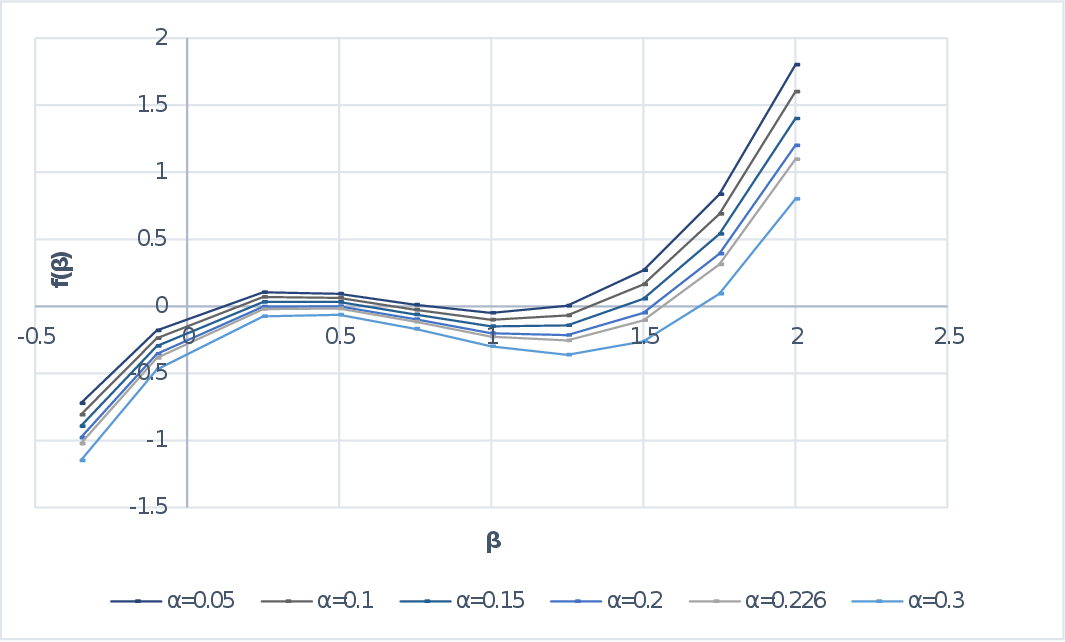}
\caption{\label{fig:cubic}Cubic equation specifying fraction of blocked txns.}
\end{center}
\end{figure}

The cubic equation as shown in Fig. \ref{fig:cubic} 
has a solution $\beta < 1$ for $\alpha \leq \alpha^* = 0.226$ and $\beta^* = 0.378$
The system becomes susceptible to thrashing for $\alpha > \alpha^*$,
but attains a maximum throughput at $\beta \approx 0.3$.

When the per step processing times are different in addition to $\alpha$
we need the fraction of lock conflicts with blocked txns $\rho=\bar{L}_b/\bar{L}$,
where $\bar{L} = \bar{L}_a + \bar{L}_b$,
where $\bar{L}_a$ is the average number of locks held by active txns \cite{Thom91,Thom93}.
The mean waiting time due to txn blocking $W$) is expressible 
as the mean waiting tine with respect to an active txn $W_1$.
Multiplying both sides by $K_1 p_c R(M)$ yields:

\begin{eqnarray}
W = W_1 \left[ 1  + \frac{ \rho(1+\rho)} {2(1-\rho^2)} \right],
\hspace{5mm}
\beta = \alpha \left[ 1 + \frac{ \rho (1+\rho) } {2(1-\rho^2)} \right]
\end{eqnarray}

Iteration is used until convergence is attained with respect to $W$.
At peak throughput $0.2 \leq \rho \leq 0.3$ which corresponds to $1.25 \leq \mbox{conflict-ratio} \leq 1.43$. 
The {\it Conflict Ratio - CR} is the ratio of the total number of locks held by all txns 
and those held by blocked txns or $\bar{L}/\bar{L_b}$ \cite{MoWe91},\cite{WHMZ94}, 
It is easy to show that $\rho= 1- 1/\mbox{CR}$,

This analysis was extended to HDAM in \cite{Thom94,Thom96},
but is not repeated here for the sake of brevity.
$\mbox{conflict-ratio}=1/(1-\rho)$ was observed in experiments to be in the range (1.26-1.60) 
which is consistent with \cite{WHMZ94}.

\section{Methods to Reduce Lock Contention}\label{sec:loadcontrol}

\hspace{3mm}
{\bf Wait-Die - WD \& Wound Wait - WW methods:} 
These two methods were proposed in the context  of distributed databases in \cite{RoSL78}.
Consider txn $T_i$ which requests a lock held by $T_j$.
In the case of WD if $T_i$ has a higher priority than $T_j$, i.e., it is older,
it is allowed to wait, otherwise it is aborted.
In the case of WW if $T_i$ has the higher priority than $T_j$ then $T_j$ is aborted, 
otherwise $T_i$ waits.

{\bf No Waiting - NW Method:}
The NW or Immediate Restart (upon lock conflict) policy was 
an exercise in analytic modeling of CC methods discussed in \cite{TaSG85}.
It is surprisingly one of three methods considered in \cite{AgCL87}. 
A generalization of NW aborts  a txn after a certain number of attempts following delays \cite{ChGM83}.

{\bf Cautious Waiting - CW Method:} 
Txn $T_C$ is aborted when it encounters a lock conflicts with a blocked txn $T_B$ \cite{HsZh92}.
$$T_C \rightarrow T_B \rightarrow T_A$$ 
$T_C$ is restarted after a  delay or by applying restart waiting, 
i.e., their restart is delayed until conflicting txns complete and release their locks \cite{ThRy91},

{\bf Running Priority - RP Method:} 
A blocked txn $T_B \rightarrow T_A$ blocked by $T_A$ 
when it causes a lock conflict with an active txn $T_C$ \cite{FrRo85}.
In symmetric RP an active txn $T_B$ blocking another txn ($T_C \rightarrow T_B$) is aborted 
when it itself is blocked ($T_C \rightarrow T_B \rightarrow T_A$ \cite{FrRT91,FrRT92}, 
thus deadlocks qre prevented.
At the cost of wasted processing due to txn aborts and restarts 
RP can attain a higher txn throughput than 2PL with blocking.
RP was analyzed as part of a unified framework for 
restart-oriented locking methods in \cite{Thom92b,Thom98b}.
The performance of restart-oriented locking methods was compared using simulation in \cite{Thom97}.

{\bf Adaptive Method:} 
The {\it conflict-ratio - CR} which is defined as the ratio 
of mean number of locks held by txns ($\bar{L}$)
and locks held by blocked txns ($\bar{L}_b$), so that 
$\mbox{CR} = \bar{L}/\bar{L}_b$ defined in \cite{MoWe91,MoWe92}.  
When $\mbox{CR} \geq 1.3$ the admission of new txns is suspended.

{\bf Optimum DMP Method:} 
If the number of executing txns equals to the optimum DMP suspend the admission of new txns \cite{MoWe92}.

{\bf Half-and-Half Method:} 
This method was developed by experimenting with the simulation model in \cite{AgCL87},
which compares the performance of standard locking, no waiting or immediate restart and OCC die policies..
When the fraction of blocked mature txns which have acquired 25\% of their locks exceeds 0.5:
it suspends the admission of new txns and cancel one or more blocked txns.

{\bf Feedback Method:} 
Two algorithms for adaptive adjustment of an upper bound for the concurrency level 
are proposed and compared via simulation in \cite{HeWa91,HeWa91}.
The incremental steps algorithm increases the DMP up to the point the performance degrades.
In fact according to simulation studies of txns with equal step-lengths $\alpha^* = 0.226$ in \cite{Thom93} 
has shown that the onset of thrashing takes many txn completions,
but it is attained sooner for txns with higher variability in size. 
The second algorithm is based on a parabola approximation of the performance function
(throughout versus DMP): $P(n)=a_0 + a_1 n +a_2 n^2$.
Parabola's maximum is designated the load threshold $n^* = -a_1/(2a_2)$ if $a_2 < 0$.
If the throughput has increased in the last interval increase DMP and vice-versa.

{\bf Critical Data Contention Load:}
For txns with equal step sizes DMP reaches a critical value $\alpha^* =0.226$,
beyond which the system is susceptible to thrashing.
The system throughput reaches its peak throughout at $\beta approx 0.3$,
which can be monitored for load control \cite{Thom91,Thom92,Thom93}

{\bf Txns with unequal per step processing times:}
The analysis in \cite{Thom91} of dynamic locking with equal per step processing times 
was extended to  unequal processing in \cite{Thom92,Thom93}.
The analysis requires an additional parameter $\rho$,
which is the fraction of lock conflicts with blocked txns
At peak throughput regardless of txn size distribution $0.2 \leq \rho \leq 0.3$.
$\rho$ is related to the conflict-ratio which was noted earlier.

{\bf Wait-Depth Limited - WDL Method:} 
Similarly to the RP method WDL aborts blocked txn blocking active txns,
but WDL takes into account txn progress based on a count of acquired locks \cite{FrRT91,FrRT92},
thus an active txn may be aborted if it has made little progress in $T_c \rightarrow T_B \rightarrow T_A$.
It outperforms all other known methods in simulation and analytic studies.
It can be implemented by modifying the {\it Internal Resource Lock Manager - IRLM} 
in the case of IBM's MVS OS \cite{Buze78} which is now z/OS. 

{\bf Table-driven Load Control for HDAM:}
The tables in \cite{Thom96}  provide acceptable peaks for compositions of dominant txn classes
and can be used for setting up DMPs for txn processing.
For two txn classes competing for the X-locks of a DBR a limit may be set on the sum of their DMPs.
More complex examples are given in \cite{Thom96}. 

{\bf Multiphase Txn Processing Methods:}
These methods take advantage of access invariance and buffer retention
(that recently accessed objects remain in the buffer)
so that disk accesses may not be required if a txn is restarted \cite{FrRT90,FrRT92}. 
In an extreme case virtual execution where locks are not acquired is used to run txns 
to completion without applying CC in the first phase to prime the buffer \cite{Reut85}, 
Thus lock contention is reduced at the cost of extra CPU processing. 

OCC method as described in \cite{FrRT90,FrRT92} and
Section 7.6.1 p. 568 in \cite{RaGe02} is relevant to two-phase txn processing.
The OCC die or silent commit policy executes txns to the end prefetching all database objects and  
if the txn fails validation or certification it is restarted with a primed buffer
Validation ensures that none of the objects accessed by the txn has been updated, since they were accessed.

The txn kill or broadcast commit option aborts and restart a conflicted txn 
when the txn interacts with with the CC manager as part of each database access. 
This option is appropriate for the second phase of txn execution,
but it can be substituted with lock preclaiming or static locking since the identity of locks is known.
OCC methods are analyzed in \cite{ThRy85},\cite{RyTh87}.
The effectiveness of low-cost checkpoints in main memory, called savepoints \cite{GrRe93},\cite{BeNe09}
and analyzed in the context of OCC kill policy in \cite{Thom95}.


{\bf Txn Chopping} reduces the level of lock contention 
by breaking down large txns to small txns \cite{SLSV95}. 

{\bf QURO - Query-Aware Compiler:} 
Contentious queries in txn code are automatically reordered 
such that they are issued as late as possible
QURO-generated code increases txn throughput by up to 6.53X, 
while reduce txn latency by up to 85\% \cite{YaCh16}.

{\bf CC in Multicore Systems:}
Seven CC methods given in Table 2 were simulated
in a main memory based database scaled to 1024 cores is reported in  \cite{YBP+14}.
All methods failed for different reasons,
but the choice of some methods was poor, for example NW but not WDL was considered.
There is a call for a completely  redesigned  DBMS  architecture.

{\bf Speedy Txns in Multicore in Memory Databases:}
Silo's OCC based commit protocol provides serializability 
by correct logging and recovery by linking periodically-updated epochs with the commit protocol. 
Silo achieves 700K TPS on a TPC-C workload mix on a 32-core machine \cite{TZL+13}. \newline
https://dspace.mit.edu/handle/1721.1/90269 

{\bf Two Optimizations for Multicore Txns:} 
commit-time updates and timestamp  splitting,  
can  dramatically  improve  the  high-contention performance of both OCC and MVCC. 
When combined, they lead to performance gains of 3.4X for MVCC and 3.6X for OCC in a TPC-C workload \cite{HQK+22}.

{\bf Elasticity for OLTP:}
Txn throughput in the presence of load spikes can be maintained
by providing  elasticity for distributed scale-out OLTP engines.
such as E-Store \cite{Taft14} and SQLVM  providing relational database-as-a-service \cite{Nara13}.
Public cloud vendors like Amazon AWS make it easy to scale up, rather than scale-out,
by choosing a different instance type \cite{Mirz19}.

{\bf Microservices and Microservice Specific Databases:} \newline
https://www.theregister.com/2022/11/04/microservices\_forcing\_a\_fast\_data


\section{Conclusions}\label{sec:conclusion}

Given the SLAs associated with OLTP systems it is important 
to ensure that their performance is satisfactory.
In addition to hardware resource contention,
software resource contention is also important.
Both latch and lock contention are important

We have outlined a methodology to study the effect of increased txn arrival rate 
or increased degree of concurrency on database lock contention.
This is important since in a poorly designed system such 
as early editions of IBM's DB2 with page level locking 
constrained the maximum throughout attainable by IBM mainframes circa 1990s
(private communication from IBM Fellow Dr Rick Baum,  1990).

The RP and WDL methods have been shown to attain higher txn throughputs
than strict 2PL at the cost of extra CPU processing.
Restarted txns are expected to find the data required for their processing in the database buffer, 
a feature not taken into account in some studies of concurrency control methods.
WDL applied to HDAM should be applied over multiple txn classes accessing a high contention DBR. 
Only locks accessing that DBR should be taken into account in deciding which txn to abort.

Measurements from high performance txn processing systems to gain insight into their
in developing analytic models for concurrency control.
Analysis with measured parameters should then be validated against higher level performance measures, 
such as response times.
The use of database logs in \cite{MCJM13} was an ambitious study 
to obtain HDAM parameters twenty years after the publication.
The author hopes that more extensive studies of OLTP access patterns 
will be carried out to develop performance prediction tools 
taking into account hardware and data (lock and latch) contention.

Of interest are CC mechanisms for post-relational databases, 
which include multidimensional databases, column databases, 
object-relational databases, NoSQL, flat file databases and others. 

\section*{Appendix: Measuring Db2 Locks and Latches}

IBM's Db2 combines many serialization mechanisms.
(1) Db2 latches, 
(2) {\it Internal Resource Lock Manager - IRLM} latches, 
(3) Locks, 
(4) Global locks for data sharing environment \cite{Otta18}.

In the case of Db2 latches the rule of thumb about the rates is:
(1) $\mbox{rate} < 1000$ no problem,  
(2) $1000 < \mbox{rate} < 10,000$,
(3) $\mbox{rate} > 10,000 $ investigate reason.

Latches are preferable to locks in terms of performance and concurrency,
since they bypass the IRLM, which manages lock requests. 
The rule of thumb is that IRLM latch suspensions should not exceed 5\% of request
Tracking locking rate it over time allows intercepting anomalies and changes in the Db2 workloads. 

Statistics on deadlock resolution victims 
are gathered with certain SMF options enabled. 
Only deadlocK victim are identified,
but more detail are obtained by enabling class 3 statistics in SMF.

Lock metrics at varying isolation levels are collected:
(1) Repeatable Read (RR), 
(2) Read Stability (RS), 
(3) Cursor Stability (CS), and 
(4) Uncommitted Read (UR) \cite{GrRe93}.

Lock escalation replaces a large number of page locks in a table with a single table lock,
which in the case of S- or X-lock escalation may lead to deadlocks.
A timeout occurs when a unit of work suspension exceeds timeout value.


\end{document}